\documentclass[prl, preprintnumbers, twocolumn, amsmath, amssymb, superscriptaddress]{revtex4}
\usepackage{graphicx}
\usepackage{dcolumn}
\usepackage{bm}

\begin{document}

\title{Tuning of Exciton States in a Magnetic Quantum Ring}
\author{Areg Ghazaryan}
\affiliation{Department of Physics and Astronomy, University of
Manitoba, Winnipeg, Canada R3T 2N2}
\author{Aram Manaselyan}
\affiliation{Department of Solid State Physics, Yerevan State
University, Yerevan, Armenia}
 \author{Tapash Chakraborty}
\affiliation{Department of Physics and Astronomy, University of
Manitoba, Winnipeg, Canada R3T 2N2}

\begin{abstract}
We have studied the exciton states in a CdTe quantum ring in an external magnetic field
containing a single magnetic impurity. We have used the multiband approximation
which includes the heavy hole - light hole coupling effects. The electron-hole spin
interactions and the $s,p-d$ interactions between the electron, hole and the magnetic
impurity are also included. The exciton energy levels and optical transitions are
evaluated using the exact diagonalization scheme. We show that due to the spin interactions
it is possible to change the bright exciton state into the dark state and vice versa
with the help of a magnetic field. We propose a new route to experimentally estimate
the $s, p-d$ spin interaction constants.

\end{abstract}

\maketitle

Electronic properties of planar nanoscale semiconductor structures, such as quantum rings 
(QRs) \cite{Qrings} and quantum dots (QDs) \cite{Qdots} have enjoyed widespread attention 
in the past few decades due to their novel fundamental effects and for potential 
technological applications. Experimental advances in creating these structures from a 
two-dimensional electron gas by using suitable confinements have resulted in confirmation 
of several theoretical predictions in these systems \cite{QR_expt,heitmann}. It has been
realized lately that QD doped with a single magnetic impurity \cite{Reiter,Besombes2005,ImpQD}
has great potential to contribute significantly in the burgeoning field of single spin
manipulation \cite{hanson}, which will eventually lead to important contributions in quantum 
information processing. Quite naturally, quantum dots, in particular the CdTe QDs containing 
a single Mn atom has been widely studied in the literature \cite{cdte_QD}. It has been 
proposed that magnetic doping of QDs provides an interesting route to magnetism in the QDs 
that can be tuned \cite{ramin}. Against the backdrop of these important developments, no such
studies involving a quantum ring have been reported yet in the literature. Recently, CdTe QRs 
have been realized experimentally \cite{Park}. Here we report on our studies of the exciton 
states in a CdTe QR in a magnetic field, containing a single magnetic impurity. We have found 
that, due to the resulting spin interactions the bright exciton state can be changed to the 
dark state and vice versa, with the help of an applied magnetic field. Additionally, we 
propose here an experimental means to estimate the $s, p-d$ spin interaction constants.

We study the exciton states in a CdTe quantum ring containing a single manganese magnetic
impurity (Mn) and subjected to a perpendicular magnetic field. Usually, the thickness
of the ring is smaller than the radial dimensions. Therefore, our system can be considered
as quasi two-dimensional, with internal radius $R^{}_1$ and the external radius $R^{}_2$.
The electron and hole are always in the ground state for the $z$ direction. We chose
the confinement potential of the quantum ring in the radial direction with infinitely high
borders: $V_{\rm conf}(\rho)=0$ if $R^{}_1\leq \rho \leq R^{}_2$ and infinity outside of the 
QR. The Hamiltonian of the system can then be written as
\begin{equation}\label{Ham}
{\cal H}={\cal H}^{}_{\rm e} + {\cal H}^{}_h+V^{}_{\rm eh}+{\cal H}^{}_{\rm eh}+{\cal
H}^{}_{\rm s-d}+{\cal H}^{}_{p-d}+{\cal H}^{}_{\text{Mn}},
\end{equation}
where ${\cal H}^{}_{\rm s-d}=-J^{}_{\rm e} \delta({\bf r}^{}_{\rm e}-{\bf
r}^{}_{\text{Mn}})\boldsymbol{\sigma} {\bf S}$ and ${\cal H}^{}_{\rm p-d}=-J^{}_{\rm h}
\delta({\bf r}^{}_{\rm h}-{\bf r}^{}_{\text{Mn}}){\bf j} {\bf S}$ describe the
electron-Mn and hole-Mn spin-spin exchange interaction with strengths $J^{}_{\rm e}$ and
$J^{}_{\rm h}$ respectively, $\bf {r}^{}_{\rm Mn}$ is the radius vector of the Mn atom.
${\cal H}^{}_{\rm eh}=-J^{}_{\rm eh} \delta({\bf r}^{}_{\rm e}-{\bf r}^{}_{\rm h})
\boldsymbol{\sigma} \bf{j}$ is the electron-hole spin
interaction Hamiltonian \cite{Efros}. The Coulomb interaction between electron and hole
term is $V^{}_{\rm eh}=-e^2/\varepsilon |\bf{r}^{}_{\rm e}-\bf{r}^{}_{\rm h}|$,
where $\varepsilon$ is the dielectric constant of the system. The last term in Eq.~(\ref{Ham})
is the Zeeman splitting for the impurity spin.

The electron Hamiltonian in our system is
\begin{equation}\label{eHam}
{\cal H}^{}_{\rm e}={\frac1{2m^{}_{\rm e}}} \left({\bf p}-{\frac e c} {\bf A}\right)^2 +
V_{\rm conf} (\rho, z) + \tfrac12 g^{}_{\rm e} \mu^{}_B B \sigma^{}_z,
\end{equation}
where ${\bf A}=\frac12B(-y,x,0)$ is the symmetric gauge vector potential and the last term is 
the electron Zeeman energy. Without the magnetic field the eigenfunctions of ${\cal H}^{}_{\rm
 e}$ can be cast in the form
\begin{equation}\label{eBazis}
\psi_{nl\sigma}^{\rm e}(\rho,\varphi)=C^{}_{nl}\,
{\rm e}^{{\rm i}l\varphi}\,f^{}_{nl}(\rho)\chi^{}_\sigma, \end{equation} where
$C^{}_{nl}$ is the normalization constant, $n=1,2,...,$ and
$l=0,\pm1,\pm2,...$ are the radial and angular quantum numbers
respectively, $\sigma$ is the electron spin and $\chi^{}_\sigma$ is the electron spin wave 
function. The functions $f^{}_{nl}(\rho)$ are obtained from a suitable linear combination of 
the Bessel functions
\begin{equation}\label{fnl}
f^{}_{nl}(\rho)=J^{}_l(k^{}_{nl}\rho)-\frac{J^{}_l(k^{}_{nl}R^{}_1)}{Y^{}_l(k^{}_{nl}R^{}_1)}
Y^{}_l (k^{}_{nl}\rho),
\end{equation}
where $k^{}_{nl}=\sqrt{2m^{}_{\rm e}E^{}_{nl}/\hbar^2}$. The corresponding eigenvalues 
$E^{}_{nl}$ are obtained from the standard boundary conditions of the eigenfunctions.

Taking into account only the $\Gamma^{}_8$ states which correspond to the states with the hole
spin $j=3/2$ and include the heavy hole - light hole coupling effects, we construct the 
single-hole Hamiltonian for the ring as
\begin{equation}\label{hHam}
{\cal H}^{}_{\rm h}={\cal H}^{}_{\rm L} + V^{}_{\rm conf}(\rho)-2\kappa \mu^{}_B B j^{}_z.
\end{equation}
Here ${\cal H}^{}_{\rm L}$ is the Luttinger hamiltonian in axial representation obtained with 
the four-band \textbf{k$\cdot$p} theory \cite{Lutt2,Pedersen2}
\begin{equation}\label{HLut}
{\cal H}^{}_{\rm L}=\frac1{2m^{}_0}\left( \begin{array}{cccc}
{\cal H}^{}_{\rm h} & R & S & 0 \\
R^* & {\cal H}^{}_l & 0 & S \\
S^* & 0 & {\cal H}^{}_l & -R \\
0 & S^* & -R^* & {\cal H}^{}_{\rm h}
\end{array}\right),
\end{equation}
where
\begin{eqnarray*}
{\cal H}^{}_{\rm h} & = & (\gamma^{}_1 + \gamma^{}_2)(\Pi_x^2 + \Pi_y^2)+
(\gamma^{}_1-2\gamma^{}_2)\Pi_z^2,\\
{\cal H}_l & = & (\gamma^{}_1-\gamma^{}_2)(\Pi_x^2 + \Pi_y^2)+
(\gamma^{}_1 + 2\gamma^{}_2)\Pi_z^2,
\end{eqnarray*}
$R=2\sqrt3\gamma^{}_3{\rm i}\Pi_-\Pi^{}_z, \quad S=\sqrt3\gamma \Pi_-^2,
\quad \gamma=\frac12(\gamma^{}_2+\gamma^{}_3),$ and ${\bf \Pi}={\bf
p}-\frac{e}{c}{\bf A}, \quad \Pi^{}_\pm=\Pi^{}_x\pm{\rm i}\Pi^{}_y.$
$\gamma^{}_1, \gamma^{}_2$, $\gamma^{}_3$ and $\kappa$ are the Luttinger
parameters and $m^{}_0$ is the free electron mass.

The Hamiltonian (\ref{hHam}) is rotationally invariant. Therefore it will be useful to 
introduce the total momentum ${\bf F}={\bf j}+{\bf l}^{}_{\rm h}$, where ${\bf j}$ is the 
angular momentum of the band edge Bloch function, and ${\bf l}^{}_{\rm h}$ is the envelop 
angular momentum. Since the projection of the total momentum $F^{}_z$ is a constant of motion,
we can find simultaneous eigenstates of (\ref{hHam}) and $F^{}_z$ \cite{Sersel}.

For a given value of $F^{}_z$ it is logical to seek the eigenfunctions of the Hamiltonian
(\ref{hHam}) as an expansion \cite{Pedersen2,Manaselyan}
\begin{equation}\label{hBazis}
\Psi^{}_{F^{}_z}(\rho,\varphi)=\sum_{n,j^{}_z}C^{}_{F^{}_z}(n,j^{}_z)
f_{n,F^{}_z-j^{}_z}^h(\rho){\rm e}^{i(F^{}_z-j^{}_z)\varphi}\chi^{}_{j^{}_z},
\end{equation}
where $\chi^{}_{j^{}_z}$ are the hole spin functions and $f_{nl}^h(\rho)$ are the radial wave
functions similar to (\ref{fnl}) with $k_{nl}^h=\sqrt{2m^{}_0E^{}_{nl}/\hbar^2(\gamma^{}_1+
\gamma^{}_2)}$. All single hole energy levels and the expansion coefficients are evaluated
numerically using the exact diagonalization scheme \cite{Manaselyan}.

In order to evaluate the energy spectrum of the exciton system we need to diagonalize the
Hamiltonian (\ref{Ham}) without spin interactions in a basis constructed as products of the
single-electron and single-hole wave functions. The good quantum number is the projection
$M^{}_z$ of the exciton total momentum $\bf{M}=\bf{F}+\bf{l}^{}_{\rm e}$. For a given value of
$M^{}_z$ and the electron spin $\sigma$ the exciton wave function can be presented as
\begin{equation}\label{exBazis}
\Psi^{}_{M^{}_z\sigma}=\sum_{n^{}_el^{}_{\rm e}}\sum_{F^{}_z}
C(n_{\rm e},l_{\rm e},F_z)\psi_{n^{}_{\rm e}l^{}_{\rm e}\sigma}^e(\rho^{}_e,
\varphi^{}_{\rm e})\Psi^{}_{F^{}_z}(\rho^{}_{\rm h}, \varphi^{}_{\rm h})
\end{equation}
The numerical calculations were carried out for a CdTe quantum ring with sizes $R^{}_1 =
100$\AA, $R^{}_2 = 300$\AA, $L^{}_z = 30$\AA\ and with the following parameters: 
$E^{}_{\rm g} = 1.568$ eV, $m^{}_{\rm e} = 0.096 m^{}_0$, $g^{}_{\rm e} = -1.5$ $\gamma^{}_1
= 5.3$, $\gamma^{}_2 = 1.7$, $\gamma^{}_3=2$, $\kappa=0.7$ \cite{Adachi}.

To include the spin-spin interactions, we can construct the wave function of the exciton and 
the magnetic impurity as an expansion of the direct products of the lowest state exciton wave 
function (\ref{exBazis}) and eigenfunctions for the magnetic impurity.
\begin{equation}\label{Bazis}
\Psi=\sum_\sigma \sum_{M^{}_z} \sum_{S^{}_z}
C(\sigma,M^{}_z,S^{}_z)\Psi^{}_{M^{}_z,\sigma}\times |S^{}_z\rangle.
\end{equation}
Here $\sigma=\pm1/2$, $S^{}_z=\pm1/2, \pm3/2, \pm5/2$ and $M^{}_z=\pm1/2, \pm3/2,\pm5/2 
\ldots$. Using the components of this expansion as the new basis functions we can calculate 
the corresponding matrix elements for the electron-hole, the electron-impurity and the 
hole-impurity interactions. Employing the steps used in \cite{ImpQD} for the electron-hole 
spin interaction matrix element, we get
\begin{equation}\label{Meh}
M^{}_{\rm eh}=-J^{}_{\rm eh} \delta^{}_{S^{}_z,S_z'}\sum_{j^{}_z,j_z'}A^{}_{\rm eh}(j^{}_z,
j_z') \left\langle\sigma,j^{}_z|\boldsymbol{\sigma} {\bf j}|\sigma',j_z'\right\rangle,
\end{equation}
where $A^{}_{\rm eh}$ is obtained by the integration of the electron and hole coordinate wave
functions, $\boldsymbol{\sigma}$ is the Pauli spin operator and $\bf{j}$ is the hole spin 
operator \cite{ImpQD}.

In the case of the electron-impurity interaction we get
\begin{gather}\label{Msd}
M^{}_{\rm s-d}=-J^{}_{\rm e}\sum_{l^{}_{\rm e},l_{\rm e}'}\delta^{}_{M^{}_z-l^{}_{\rm e},M_z'-
l_{\rm e}'}A^{}_{\rm s-d}({\bf r}^{}_{\rm e}={\bf r}^{}_{Mn},l^{}_{\rm e},l_{\rm e}')
\times \nonumber \\
\left\langle\sigma^{}_z,S^{}_z| \boldsymbol{\sigma}{\bf{S}}|\sigma_z',S_z'
\right\rangle,
\end{gather}
where $A^{}_{\rm s-d}$ is obtained after the integration of the hole coordinate wave functions
and putting ${\bf r}^{}_{\rm e}={\bf r}^{}_{\rm Mn}$ in the electron wave function.
Similarly, for the case of hole-impurity interaction we get
\begin{gather}\label{Mpd}
M^{}_{\rm p-d}=-J^{}_{\rm h} \delta^{}_{\sigma,\sigma'}\sum_{j^{}_z,j_z'} A^{}_{\rm p-d}({\bf
r}^{}_{\rm h}={\bf r}^{}_{\rm Mn},j^{}_z,j_z')
\times \nonumber\\
\left\langle j^{}_z,S^{}_z|{\bf{j}}{\bf{S}}|j_z',S_z' \right\rangle.
\end{gather}
In order to calculate the spin matrix elements we need to introduce the raising
and lowering operators
\begin{eqnarray}
{\bf{S^{}_+}}|S^{}_z\rangle=\sqrt{S(S+1)-S^{}_z(S^{}_z+1)}|S^{}_z+1\rangle,
\nonumber\\
{\bf{S^{}_-}}|S^{}_z\rangle=\sqrt{S(S+1)-S^{}_z(S^{}_z-1)}|S^{}_z-1\rangle.
\end{eqnarray}
As the spin interactions are short ranged, the most interesting case is when the magnetic
impurity is located in the region of average ring radius. In that case we can take 
$\rho^{}_{\rm Mn} = (R^{}_1 + R^{}_2)/2$ and $\varphi^{}_{\rm Mn} = 0$. The problem was 
solved numerically using the exact diagonalization scheme and with interaction parameters 
$J^{}_{\rm e} = 15$ meV nm$^3$, $J^{}_{\rm h} = -60$ meV nm$^3$ \cite{Reiter,Besombes2005}.

In order to evaluate the optical transition probabilities, let us note that the initial state
of the system is that of the magnetic impurity spin with the valence band states fully 
occupied and the conduction band states being empty. Let us also assume that the impurity 
states are pure coherent states $|{\rm i}\rangle=|S^{}_z\rangle$. Recently there were several 
experimental reports where the quantum dots with a single magnetic impurity in a coherent 
spin state were prepared even in the absence of a magnetic field \cite{Reiter,Besombes2009}. 
The final states are the eigenstates of the Hamiltonian (\ref{Ham}) presented in (\ref{Bazis})
$|f\rangle=|\Psi \rangle$. In the electric dipole approximation the relative oscillator 
strengths for all possible optical transitions are proportional to $P(m)\sim\left|\langle
\Psi|m,S^{}_z\rangle \right|^2.$ Here the values of $m = 1,0,-1$ characterize the polarization
of the light as $\sigma^+$, $\pi$ and $\sigma^-$ respectively \cite{Bhattacharjee2003}. It 
should also be mentioned that the impurity spin state remains unchanged during the optical 
transitions.

In the absence of the magnetic atom in the QR and without the electron-hole spin interaction, 
the ground state of the exciton will be four-fold degenerate with values of the total momentum
$\pm1$ and $\pm2$. The magnetic field lifts that degeneracy due to the Zeeman splitting and as 
a result two bright ($J^{}_z=\pm1$) and two dark ($J^{}_z=\pm2$) exciton states appear. The 
electron-hole spin exchange interaction in turn gives rise to a further splitting between the 
bright and dark exciton states and removes the degeneracy between them in zero magnetic field. 
In Fig.~1 (a) the dependence of few low-lying exciton energy levels on the magnetic field is 
presented with the electron-hole spin interaction included, for the QR without a magnetic 
impurity. The corresponding optical transition probabilities for $\sigma^-$ and $\sigma^+$ 
polarizations are shown in Fig.~1 (b). The sizes of the symbols in Fig.~1 (b) indicate the 
probability of the optical transition to that state. For smaller values of the magnetic
field, two lowest energy levels in Fig.~1 (a) correspond to the dark exciton states and 
hence the transition probabilities to that states are very weak. The energies of two bright 
exciton states with the most important components of the basis functions $|\sigma,j^{}_z
\rangle=|-1/2,3/2\rangle$ and $|1/2,-3/2\rangle$ are shifted upwards by the electron-hole
spin interaction, but still are clearly visible optically in Fig.~1 (b). In the case of 
$\sigma^+$ polarization we have a strong transition to the state $|-1/2,3/2\rangle$ (black 
squares), and for the case of $\sigma^-$ polarization, the strong transition is for the state 
$|1/2,-3/2\rangle$ (white squares). It should be also mentioned that with the increase of the 
magnetic field the transition probabilities remain almost unchanged.

\begin{figure}
\begin{center}\includegraphics[width=5cm]{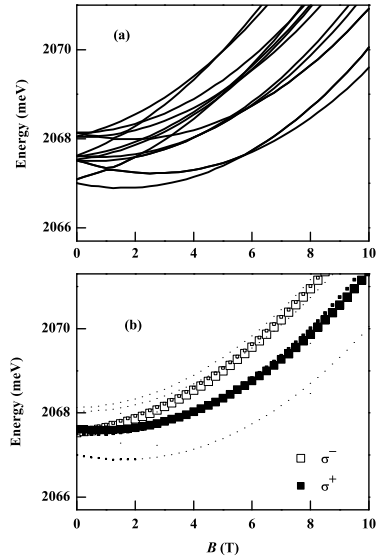}\end{center}
\caption{\label{fig:Exciton} (a) Magnetic field dependence of the exciton energy levels with
electron hole spin-interaction included. (b) Optical transition amplitudes for the $\sigma^+$
and $\sigma^-$ polarizations.}
\label{fig1}
\end{figure}

The Mn atom has a spin $S=5/2$ and there are six possible values of the impurity spin 
projection $S^{}_z$. That is why due to the $s, p-d$ spin interaction each exciton energy 
level presented in Fig.~1 (a) will split into six. In our calculations we consider the 
energies of first twelve lowest exciton states therefore there are 72 energy levels presented. 
Due to Zeeman splitting and $s, p-d$ splitting of energy levels there will be many level 
crossings and anticrossings. The presence of the impurity inside the ring material removes the 
symmetry of the structure and now we do not have any good quantum numbers to describe the 
states. All states are mixed supperpositions with different values of total momentum of 
electron, hole and magnetic impurity.

\begin{figure}
\includegraphics{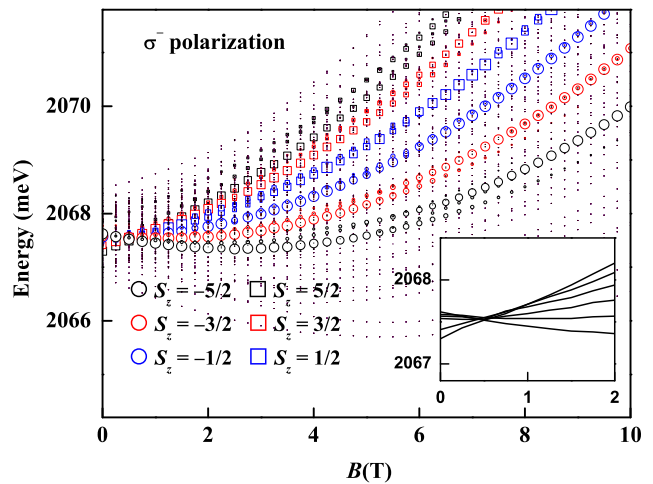}
\caption{\label{fig:Espd_2} Magnetic field dependence of the optical
transition amplitudes for the case of $\sigma^-$ polarization and
for various values of the initial state impurity spin projection
$S^{}_z$. The crossing point of six bright exciton states is shown
as inset.}
\end{figure}

In order to clarify this complicated situation, we have considered here the optical transition 
spectrum to these 72 states. As we have mentioned above the initial state is a pure coherent 
state with a fixed value of the impurity spin $S^{}_z$. The final states are the exciton states 
with the magnetic impurity. The high probability transitions will be possible only to the bright 
exciton states, which have the most important components with the same value of the
impurity spin $S^{}_z$. The results for the $\sigma^-$ polarization of the incident light are 
presented in Fig.~2. Here the shapes and the colors of the points indicate the initial spin of 
the impurity and the sizes of the points indicates the probability of the transition to that 
state. For the $\sigma^-$ polarization of the incident light, the bright exciton states must 
have the important component with $|\sigma^{}_z,M^{}_z\rangle=|1/2,-3/2 \rangle$. For the 
$\sigma^+$ polarization (Fig.~3) the most important component of the bright states must be 
$|-1/2,3/2\rangle$ \cite{Bhattacharjee2003}. For example in the case of the $\sigma^-$
polarization and for the initial state $S^{}_z=-5/2$ in low magnetic fields we have only one 
strong transition (Fig.~2 black circles). But near the fields of 5-6 Tesla that line weakens and 
disappears and a new optical mode appears. Similar behavior can also be seen for other impurity 
spin states. This effect is the direct signature of the $s, p-d$ spin interaction. Due to spin 
interactions now the bright exciton state $|1/2,-3/2\rangle$ is coupled with the dark state 
$|1/2,-1/2\rangle$ and we have two coupled energy levels. For the first level at $B=0$ the weight 
of the $|1/2,-3/2,-5/2\rangle$ state is 0.96 and the weight of the $|1/2,-3/2,-5/2\rangle$ state 
is 0.19. With the increase of the magnetic field the weight of $|1/2,-3/2,-5/2\rangle$ decreases 
and the weight of $|1/2,-1/2,-5/2\rangle$ state increases. As a result the bright state changes 
to dark. For the second level we have an opposite picture. In the case of $\sigma^+$ polarization 
(Fig.~3 (a)-(d)) we see similar effects for the case of $S^{}_z=\pm5/2$ and $\pm3/2$. Now the 
bright exciton state $|-1/2,3/2,S^{}_z\rangle$ is coupled with the dark state $|-1/2,5/2,S^{}_z
\rangle$. For the case of $S^{}_z=\pm1/2$ the effect is not pronounced because the energies of 
the mixed bright and dark states are too close to each other.

\begin{figure}
\includegraphics{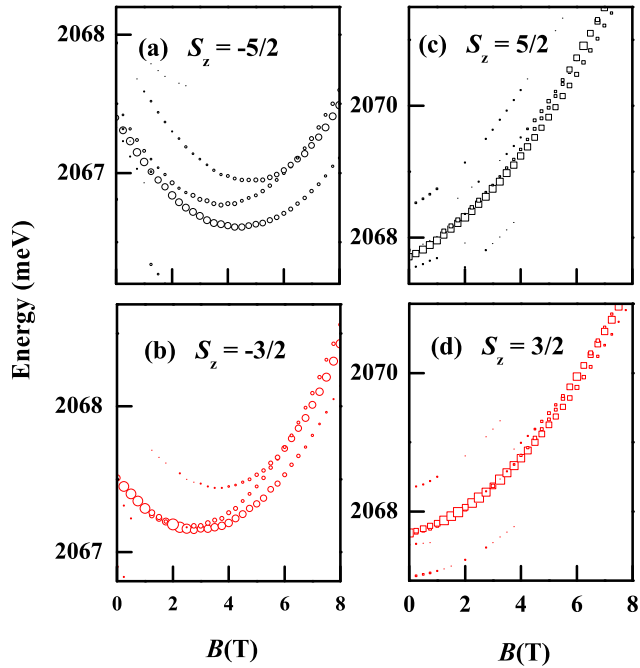}
\caption{\label{fig:Espd} Magnetic field dependence of the optical transition amplitudes for the
case of $\sigma^+$ polarization and for various values of the initial state impurity spin
projection $S_z$.} \end{figure}

We should be mention here about an interesting effect observed in the case of $\sigma^-$ 
polarization. In Fig.~2 there is a crossing point for all energies of the bright exciton
states and for $B = 0.5$ Tesla (see inset in Fig.~2). This interesting effect can be explained 
as follows: In the case of the $\sigma^-$ polarization the most important component of the 
bright exciton states is $|1/2,-3/2,S^{}_z\rangle$, where $S^{}_z$ takes six possible values. 
For all these states the energy term connected with the $s, p-d$ spin interactions has opposite 
sign with the Zeeman splitting energy of the magnetic impurity $g^{}_{\rm Mn}\mu^{}_B B S^{}_z$, 
where $g^{}_{\rm Mn}=2$. For a certain value of the magnetic field $B^{}_0$ these two terms will 
cancel each other and we will see a crossing point. In our case $B^{}_0=0.5$ Tesla, but in 
general, the value of $B^{}_0$ depends on the ring parameters and on the $s, p-d$ interaction 
constants $J^{}_{\rm e}$ and $J^{}_{\rm h}$. We believe that this effect is experimentally
observable. After the detection of the experimental value of the crossing point $B^{}_0$
one should be able to estimate the real values of the $s, p-d$ interaction constants $J^{}_{\rm 
e}$ and $J^{}_{\rm h}$ in a quantum ring. In the case of the $\sigma^+$ polarization the most 
important component of the bright states is $|-1/2,3/2,S_z\rangle$. Now the $s, p-d$ interaction 
term and the Zeeman splitting term for the magnetic impurity always have the same
sign and there is no crossing point.

In conclusion, we have studied the effect of spin interactions on the exciton states in a 
quantum ring with a single magnetic impurity subjected to a perpendicular magnetic field. The 
optical properties of such a QR have been investigated. It was shown that due to the $s, p-d$ 
spin exchange interactions between the electron, hole and the magnetic impurity it is possible
to change the bright exciton state into a dark state and vice versa with the help of the 
applied magnetic field. Additionally, a new method is proposed for experimental estimation of 
$s, p-d$ spin interaction constants.

The work has been supported by the Canada Research Chairs Program of the Government of Canada,
the Armenian State Committee of Science (Project no. SCS-13-1C196), and by the ANSEF grant 
Nano-3334.

\end{document}